\documentclass[5p,fleqn,twocolumn,final]{elsarticle}
\biboptions{sort,compress}
\usepackage{showkeys}
\usepackage{amssymb}
\usepackage[mathscr]{eucal}
\usepackage{amsmath}
\usepackage{graphicx}
\usepackage{pstricks}
\usepackage{wrapfig}
\newcommand{\e}{\mathbf}

\journal{Physics Letters A}

\begin{document}

\begin{frontmatter}

\title{Excitations and spin correlations  near the interface of \\ two three-dimensional Heisenberg antiferromagnets}

\author{N. Voropajeva and A. Sherman}

\address{Institute of Physics, University of Tartu, Riia 142, 51014 Tartu, Estonia}

\begin{abstract}

Magnetic excitations and spin correlations near the interface of two spin-$\frac12$ Heisenberg antiferromagnets are considered using the spin-wave approximation. When the interaction between boundary spins differs essentially from exchange constants inside the antiferromagnets, quasi-two-dimensional spin waves appear in the near-boundary region. They eject bulk magnons from this region, thereby dividing the antiferromagnets into areas with different magnetic excitations. The decreased dimensionality of the near-boundary modes leads to amplified nearest-neighbor spin correlations in the interface area.

\end{abstract}


\end{frontmatter}

\section{Introduction}

In recent years, an active interest has been taken in heterostructures fabricated out of magnetic crystals. Looking for new effects and their possible applications, a wide variety of systems has been investigated both experimentally and theoretically (see, e.g., \cite{ohtomo,brinkman,reyren,dikin,li}). One of the systems belonging to this group is the interface of a Heisenberg antiferromagnet with vacuum. Spin excitations and correlations near this interface were investigated in Refs.\ \cite{metlitski,hoglund,pardini,voropajeva2009,voropajeva2010,sherman}. In particular, it was shown that the antiferromagnet is divided into two regions with different spin excitations \cite{voropajeva2009,voropajeva2010,sherman}. In the first region, which embraces two near-boundary atomic layers, the excitations have dimensionality one less than the crystal dimensionality. The rest of the crystal is the second region, which elementary excitations are standing spin waves. The near-boundary excitations eject these latter spin waves from the near-boundary region, which leads to the mentioned separation of the antiferromagnet into the two regions. Both physically and mathematically this situation resembles the problem of the local impurity considered by I.M. Lifshits \cite{lifshits}, in which localized electronic states eject bulk states from the impurity region. The decreased dimensionality of the near-boundary modes leads to increased spin correlations in the interface region and their damped oscillations away from the region \cite{metlitski,hoglund,pardini}.

In this Letter the heterostructure of two semi-infinite spin-$\frac12$ Heisenberg antiferromagnets with nearest-neighbor exchange constants $J_1$ and $J_2=0.8J_1$ is considered. Such heterostructures were already fabricated \cite{satoh,wojek}; however, to our knowledge magnetic excitations near the interface were not experimentally investigated yet. Both antiferromagnets are supposed to have identical simple cubic lattices with the interface coinciding with a crystallographic plane. As in the case of the antiferromagnet-vacuum interface, for the exchange constant between sites of the two antiferromagnets $J_3\ll J_2$ modes of decreased dimensionality exist in the near-boundary region, expelling bulk modes from this area.  These near-boundary modes are seen as sharp peaks in the spectral function characterized by the layer index and the wave vector $\e k$ describing movement parallel to the interface. With $J_3$ approaching $J_2$ the near-boundary mode disappears at first in the antiferromagnet with $J=J_2$ and for small $\e k$. With further growth of $J_3$ the mode fades away also in the second antiferromagnet. A new type of near-boundary excitations appears at $J_3\approx J_1$ in the antiferromagnet with $J=J_2$, and at a somewhat larger $J_3$ in the other antiferromagnet. Their frequencies surpass frequencies of bulk spin waves. These interface excitations are mainly connected with pairs of spins on each side of the boundary. In the near-boundary region, the nearest-neighbor spin correlations are amplified in comparison with the bulk value, when $J_3\ll J_2$ and $J_3>J_1$. The deviations nearly disappear when $J_3$ approaches $J_2$. At $J_3\approx(J_1+J_2)/2$ the deviations reverse their signs.

\section{Main formulas}
The axes are chosen in such a way that the antiferromagnet with the exchange constant $J_1$ is located in the half-space $l_x\geqslant0$, while the other antiferromagnet with the exchange constant $J_2$ is in the half-space $l_x\leqslant-1$. Additionally there is an exchange interaction~$J_3$ between boundary spins of the antiferromagnets. Here sites of a three-dimensional simple cubic lattice are labeled by the three coordinates $l_x,l_y,l_z$, and the lattice spacing is set as the unit of length. The system is described by the Hamiltonian
\begin{eqnarray}\label{1}
    H&=&\frac{J_1}{2}\sum_{\e l \e a}\sum_{l_x\geqslant 0}\e S_{\e l+\e a,l_x}\e S_{\e l l_x}+J_1\!\!\!\sum_{\e l,l_x\geqslant0}\e S_{\e l,l_x+1}\e S_{\e l l_x}\nonumber\\
    &+&\frac{J_2}{2}\sum_{\e l \e a}\sum_{l_x\leqslant -1}\e S_{\e l+\e a,l_x}\e S_{\e l l_x}+J_2\!\!\!\sum_{\e l,l_x\leqslant-1}\e S_{\e l l_x}\e S_{\e l, l_x-1}\nonumber\\
    &+&J_3\sum_{\e l}\e S_{\e l 0}\e S_{\e l,-1}\,,
\end{eqnarray}
where $\e l=(l_y,l_z)$, $\e a=(\pm1,0),(0,\pm1)$ are four unit vectors, which connect nearest neighbor sites in the $YZ$ plane, and $\e S_{\e L}$ is the spin-$\frac{1}{2}$ operator.

Since for low temperatures the system has the long-range antiferromagnetic order, its low-lying elementary excitations can be described in the spin wave approximation,
\begin{equation}\label{2}
    S^z_{\e{L}}=e^{i\e{\Pi
    L}}\left(\frac12-b^\dag_\e{L}b_{\e{L}}\right), \quad
    S^\pm_\e{L}=P^{\pm}_\e{L}b_\e{L}+P_\e{L}^{\mp}
    b^\dag_\e{L},
\end{equation}
where the spin-wave operators $b_{\e L}$ and $b^\dag_{\e L}$ satisfy the Boson commutation relations, and
\begin{equation*}
    \e{\Pi}=(\pi,\pi,\pi),
    \quad P^\pm_\e{L}=\frac12\left(1\pm e^{i\e{\Pi L}}\right), \quad
    \e L=(l_x,l_y,l_z).
\end{equation*}
If we substitute Eq.~\eqref{2} into Eq.~\eqref{1}, drop constant terms and terms containing more than two spin-wave operators, we obtain
\begin{eqnarray}\label{3}
    H&=&H_1+H_2+H_3,\nonumber\\
    H_1&=& 3J_1\sum_{\e k l_x\geqslant0}\left(1-\frac16\delta_{l_x 0}\right)b^\dag_{\e k l_x}b_{\e k l_x}\nonumber\\
    &+&J_1\sum_{\e k l_x\geqslant0}\gamma^{(2)}_{\e k}\left(b_{\e{k}l_x}b_{\e{-k},l_x}+b^\dag_{\e k l_x}b^\dag_{\e{-k},l_x}\right)\nonumber\\
    &+&\frac{J_1} 2\sum_{\e k l_x\geqslant0}\left(b_{\e k l_x}b_{\e{-k},l_x+1}+b^\dag_{\e k l_x}b^\dag_{\e{-k},l_x+1}\right),\nonumber\\
    H_2&=&3J_2\sum_{\e k l_x\leqslant-1}\left(1-\frac16\delta_{l_x, -1}\right)b^\dag_{\e k l_x}b_{\e k l_x}\\
    &+&J_2\sum_{\e k l_x\leqslant-1}\gamma^{(2)}_{\e k}\left(b_{\e{k}l_x}b_{\e{-k},l_x}+b^\dag_{\e k l_x}b^\dag_{\e{-k},l_x}\right)\nonumber\\
    &+&\frac{J_2} 2\sum_{\e k l_x\leqslant-1}\left(b_{\e k l_x}b_{\e{-k},l_x-1}+b^\dag_{\e k l_x}b^\dag_{\e{-k},l_x-1}\right),\nonumber\\
    H_3&=&\frac{J_3}{2}\sum_{\e k}\Bigl(b^\dag_{\e k0}b_{\e k0}+b^\dag_{\e k,-1}b_{\e k,-1}\nonumber\\
    &+&b_{\e k0}b_{\e -k,-1}+b^\dag_{\e k0}b^\dag_{\e -k,-1}\Bigr).\nonumber
\end{eqnarray}
Here we took into account the translational invariance of
Hamiltonian \eqref{1} in the $YZ$ plane and used the Fourier transformation
\begin{equation*}
    b_{\e k l_x}=\frac{1}{\sqrt N}\sum_{\e l}e^{i\e {kl}}b_{\e l l_x},
\end{equation*}
where $\e k$ is a two-dimensional (2D) wave vector and $N$ is the number of sites in the periodic $YZ$ plane. In Eq.~\eqref{3},
$\gamma^{(2)}_{\e k}=\frac12[\cos(k_y)+\cos(k_z)]$.

To investigate the spectrum of elementary excitations, we introduce the two-component operator
\begin{equation*}
    \hat B_{\e k l_x}=
    \left(
      \begin{array}{c}
        b_{\e k l_x} \\
        b^\dag_{-\e k l_x} \\
      \end{array}
    \right)
\end{equation*}
and define the matrix retarded Green's function
\begin{equation}\label{4}
    \hat D(\e k t l_x l'_x)=-i\theta(t)\left\langle\left[\hat B_{\e k l_x}(t),\hat B^\dag_{\e k l'_x}\right]\right\rangle,
\end{equation}
where $\hat B_{\e k l_x}(t)=e^{i H t}\hat B_{\e k l_x}e^{-H t}$
with $H$ determined by Eq.~\eqref{3}.

To calculate Green's function \eqref{4}, we use the equation of motion
\begin{eqnarray}\label{5}
    &&\hspace{-0.9cm}i\frac{d}{dt}\hat D(\e k t l_x l'_x)=
    r_1\left[3\hat\tau_3+2\gamma^{(2)}_{\e k}\right]\hat D(\e k t l_x l'_x)\nonumber\\
    &&~+ r_2\hat\tau_1\hat D(\e k t,l_x+1,l'_x)+
    r_3\hat\tau_1\hat D(\e k t,l_x-1,l'_x)\nonumber\\
    &&+\left[\delta_{l_x 0}\frac{J_3-J_1}{2}+\delta_{l_x,-1}\frac{J_3-J_2}{2}\right]\hat\tau_3\hat D(\e k t l_x l'_x)\nonumber\\
    &&+\frac{J_3}{2}\hat\tau_1\delta_{l_x,-1}\hat D(\e k t,l_x+1,l'_x)\nonumber\\
    &&+\frac{J_3}{2}\hat\tau_1\delta_{l_x 0}\hat D(\e k t,l_x-1,l'_x)+\delta(t)\delta_{l_x l'_x}\hat\tau_3,
\end{eqnarray}
where
\begin{eqnarray*}
  r_1 &=& J_2\theta(-l_x-1)+J_1\theta(l_x), \\
  r_2 &=& \frac12\left[J_2\theta(-l_x-2)+J_1\theta(l_x)\right], \\
  r_3 &=& \frac12\left[J_2\theta(-l_x-1)+J_1\theta(l_x-1)\right],\\
  \tau_1&=&\left(
           \begin{array}{cc}
             0 & 1 \\
             -1 & 0 \\
           \end{array}
         \right),\quad
  \tau_3=\left(
             \begin{array}{cc}
               1 & 0 \\
               0 & -1 \\
             \end{array}
           \right).
\end{eqnarray*}
Equation \eqref{5} contains local perturbations proportional to $\delta_{l_x 0}$ and $\delta_{l_x,-1}$. By analogy with Lifshits' solution \cite{lifshits} Green's function \eqref{4} can be expressed through the function $\hat D^{(0)}(\e k t l_x l'_x)$ of the equation without the perturbation terms. After the Fourier transformation and some mathematical manipulations this expression reads
\begin{eqnarray}\label{6}
  &&\hat D(\e k\omega l_x l'_x)= \hat D^{(0)}(\e k\omega l_x l'_x) \nonumber\\
   &&+  \frac{J_3-J_1}{2}\hat D^{(0)}(\e k\omega l_x 0)\hat D(\e k\omega 0 l'_x)\nonumber\\
   &&+  \frac{J_3-J_2}{2}\hat D^{(0)}(\e k\omega l_x,-1)\hat D(\e k\omega,-1,l'_x)\nonumber\\
   &&+ \frac{J_3}{2} \hat D^{(0)}(\e k\omega l_x 0)\hat\tau_2\hat D(\e k\omega,-1,l'_x)\nonumber\\
   &&+\frac{J_3}{2} \hat D^{(0)}(\e k\omega l_x,-1)\hat\tau_2\hat D(\e k\omega 0 l'_x),
\end{eqnarray}
where $\hat\tau_0$ is the $2\times2$ identity matrix and
\begin{equation*}
    \tau_2=\left(
             \begin{array}{cc}
               0 & 1 \\
               1 & 0 \\
             \end{array}
           \right),
\end{equation*}
\begin{equation*}
\begin{split}
     \hat D(\e k\omega,-1,l_x)&=\Bigl[\hat\tau_0-\frac{J_3-J_2}{2}\hat D^{(0)}(\e k\omega,-1,-1)\\
     &-\frac{J^2_3}{4}\hat R\hat D^{(0)}(\e k\omega 0 0)\hat\tau_2\Bigr]^{-1}\\
     &\times \left[\hat D^{(0)}(\e k\omega,-1,l'_x)+\frac{J_3}{2}\hat R \hat D^{(0)}(\e k\omega 0 l'_x)\right],
\end{split}
\end{equation*}
\begin{equation*}
\begin{split}
    \hat D(\e k\omega 0 l'_x)&=\hat W^{-1}\Bigl[\frac{J_3}{2}\hat D^{(0)}(\e k\omega 0 0)\hat\tau_2\hat D(\e k\omega,-1,l'_x)\\
    &+\hat D^{(0)}(\e k\omega 0 l'_x)\Bigr],
\end{split}
\end{equation*}
\begin{equation*}
\begin{split}
\hat R&=\hat D^{(0)}(\e k\omega,-1,-1)\hat\tau_2\left[\hat\tau_0-\frac{J_3-J_1}{2}\hat D^{(0)}(\e k\omega 0 0)\right]^{-1}\\
\hat W&=\hat\tau_0-\frac{J_3-J_1}{2}\hat D^{(0)}(\e k\omega 0 0).
\end{split}
\end{equation*}

To calculate Green's function $\hat D^{(0)}(\e k\omega l_x l'_x)$  it is necessary to diagonalize Hamiltonian \eqref{3} without the terms proportional to $J_3$, $\delta_{l_x 0}$, and $\delta_{l_x,-1}$. This can be fulfilled using the Bogo\-liu\-bov-Tyablikov transformation \cite{tyablikov}. The obtained Green's function $\hat D^{(0)}(\e k\omega l_x l'_x)$ reads \cite{sherman}
\begin{eqnarray}\label{7}
    \hat D^{(0)}(\e k\omega l_x l'_x)&=&\hat D^{(01)}(\e k\omega l_x l'_x)+\hat D^{(02)}(\e k\omega l_x l'_x),\nonumber\\
    \hat D^{(01)}(\e k\omega l_x l'_x)&=&\theta(l_x)\theta(l'_x)\int\limits^{\pi}_0 d k_x \sin[k_x(l_x+1)]\nonumber\\
    &\times& \sin[k_x(l'_x+1)]\hat T_{\e k k_x}(J_1),\nonumber\\
    \hat D^{(02)}(\e k\omega l_x l'_x)&=&\theta(-l_x-1)\theta(-l'_x-1)\int\limits^{\pi}_0 d k_x \sin(k_x l_x)\nonumber\\
    &\times& \sin(k_x l'_x)\hat T_{\e k k_x}(J_2),
\end{eqnarray}
\begin{eqnarray*}
   \hat T_{\e k k_x}(J)&=&\frac{\hat P_{\e k k_x}(J)}{\omega-E_{\e k k_x}(J)+i\eta}-\frac{\hat Q_{\e k k_x}(J)}{\omega+E_{\e k k_x}(J)+i\eta}\\
   \hat P_{\e k k_x}(J)&=&\left(
                   \begin{array}{cc}
                     A^2_{\e k k_x}(J) & A_{\e k k_x}(J)B_{\e k k_x}(J) \\
                     A_{\e k k_x}(J)B_{\e k k_x}(J) & B^2_{\e k k_x}(J) \\
                   \end{array}
                 \right),\nonumber\\
    \hat Q_{\e k k_x}(J)&=&\left(
                   \begin{array}{cc}
                     B^2_{\e k k_x}(J) & A_{\e k k_x}(J)B_{\e k k_x}(J) \\
                     A_{\e k k_x}(J)B_{\e k k_x}(J) & A^2_{\e k k_x}(J) \\
                   \end{array}
                 \right),\nonumber
\end{eqnarray*}
\begin{eqnarray*}
    A_{\e k k_x}(J) &=& \sqrt{\frac{2}{\pi}}\frac{3J+E_{\e k k_x}(J)}{\sqrt{\left[3J+E_{\e k k_x}(J)\right]^2-\left(3J\gamma^{(3)}_{\e k k_x}\right)^2}},\\
    B_{\e k k_x}(J) &=& -\sqrt{\frac{2}{\pi}}\frac{3J\gamma^{(3)}_{\e k k_x}}{\sqrt{\left[3J+E_{\e k k_x}(J)\right]^2-\left(3J\gamma^{(3)}_{\e k k_x}\right)^2}},\nonumber\\
    E_{\e k k_x}(J)&=&3J\sqrt{1-\left(\gamma^{(3)}_{\e k k_x}\right)^2}, \nonumber
\end{eqnarray*}
where $\gamma^{(3)}_{\e k k_x}=\dfrac13[\cos(k_x)+\cos(k_y)+\cos(k_z)]$ and $\eta=+0$.

\begin{figure}[t]
\begin{center}
\includegraphics[width=0.9\linewidth]
{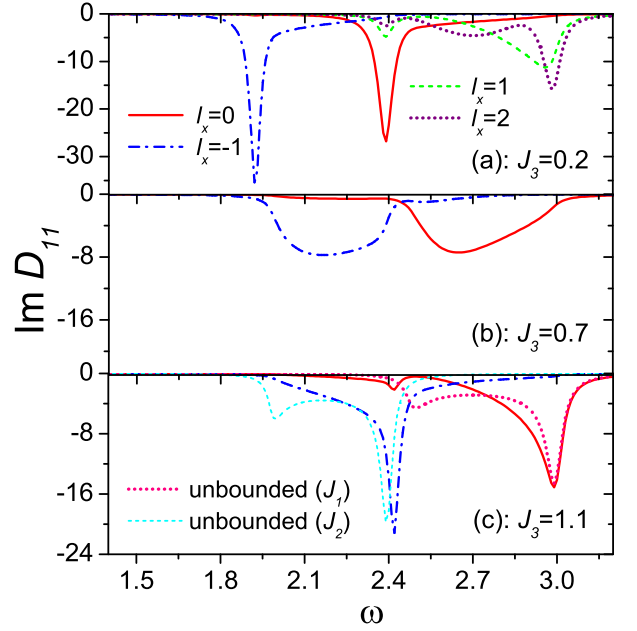} \caption{The spectral function $\mbox{Im}D_{11} ({\rm {\bf k}}\omega l_x l_x )$ for (a) $J_3 =0.2$, (b) $J_3 =0.7$ and (c) $J_3 =1.1$. ${\rm {\bf k}}=\left( {0.6\pi ,0} \right)$ and $J_2 =0.8$. Red solid and blue dash-dotted lines correspond to $l_x =0$ and $l_x =-1$, respectively. In part (a), green dashed and purple doted lines show spectral functions for $l_x =1$ and 2. In part (c), pink dotted and cyan dashed lines demonstrate spectral functions of unbounded antiferromagnets with $J=J_1 $ and $J=J_2 $, respectively.} \label{fig1}
\end{center}
\end{figure}

\section{Spin excitations}
\begin{figure}[t]
\begin{center}
\includegraphics[width=0.9\linewidth]
{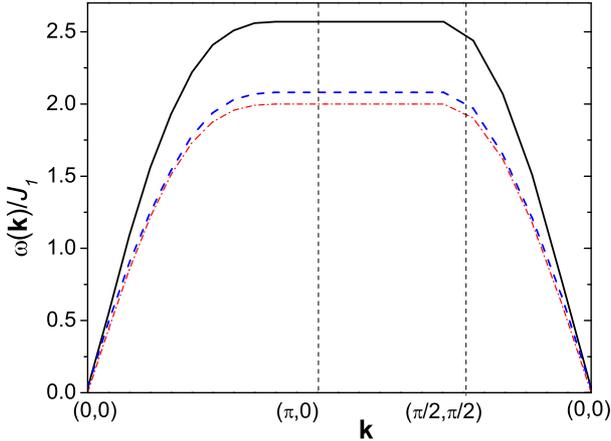} \caption{The dispersion of the near-boundary modes along the symmetry lines of the 2D Brillouin zone in the antiferromagnet with $J=J_1$ (the black solid line) and in the antiferromagnet with $J=J_2 $ (the blue dashed line). The red dash-dotted line demonstrates the dispersion of the 2D spin waves for $J=J_1 $. $J_3 =0.2$ and $J_2 =0.8$.} \label{fig2}
\end{center}
\end{figure}

Below the exchange constant $J_1$ is set as the unit of energy, and the case $J_2=0.8$  is considered. For $J_3=0$ the problem reduces to two noninteracting semi-infinite antiferromagnets, which spin excitations were considered in Refs.\ \cite {metlitski,voropajeva2009,voropajeva2010,sherman}. As mentioned in the Introduction, the spectrum of each antiferromagnet consists of a quasi-2D mode of spin waves, which are mainly located in two near-boundary layers, and bulk modes of standing spin waves in the rest of the crystal. This picture retains for $J_3\ll J_2$.  This case is illustrated in Fig.~\ref{fig1}(a). In the boundary layers $l_x=0$ and $l_x=-1$ of both semi-infinite antiferromagnets the spectral function $\mathrm{Im}D_{11}(\e k\omega l_x l_x)$ has sharp peaks, which rapidly lose intensity with distance from the interface, as seen from the spectral functions for $l_x=1$  and $2$  [the functions for $l_x=-2$ and $-3$,  which are not shown in  Fig.~\ref{fig1}(a), look analogously with the respective frequency shift]. These peaks correspond to the near-boundary modes, which propagate along the interface with the dispersion similar to that of the 2D magnons, as seen in Fig.~\ref{fig2}. In the boundary layers the continuum of standing spin waves is seen as weak shoulders on the high-frequency sides of the peaks. However, already in layers $l_x=2$ and $l_x=-3$ the peaks become weak and spectral functions become close to those in unbounded antiferromagnets shown in Fig.~\ref{fig1}(c) [maxima in these functions are connected with power divergences in $\mathrm{Im}D_{11}(\e k\omega l_x l_x)$ at the edges of the spectra $E_{\e k k_x}(J)$;
a finite artificial broadening transforms the divergences into maxima]. Thus, for $J_3\ll J_2$ the crystals are split into two regions with different spin excitations -- four layers closest to the interface are the domain of existence of the near-boundary modes with the suppressed intensity of bulk modes, while these latter modes dominate in the rest of the crystals.

\begin{figure}[t]
\begin{center}
\includegraphics[width=\linewidth]
{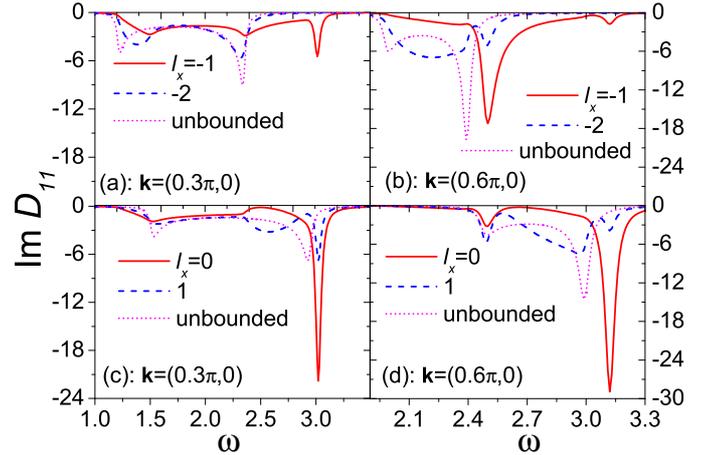} \caption{The spectral function $\mbox{Im }D_{11} ({\rm {\bf k}}\omega l_x l_x )$ for ${\rm {\bf k}}=\left( {0.3\pi ,0} \right)$ (a,c) and for ${\rm {\bf k}}=\left( {0.6\pi ,0} \right)$ (b,d). $J_3 =1.51$, $J_2 =0.8$. Red solid lines correspond to $l_x =0$ and $l_x =-1$, blue dashed lines to $l_x=1$ and $l_x =-2$ on the lower and upper panels, respectively. Pink dotted lines are spectral functions of an unbounded antiferromagnets with $J=J_1 $ and $J=J_2$.} \label{fig3}
\end{center}
\end{figure}

With increasing $J_3$ the maximum of the near-boundary mode in the spectral function weakens and disappears starting from the antiferromagnet with the smaller exchange constant and small wave vectors $\e k$. The near-boundary mode fades away at $J_3\approx0.5$ in the antiferromagnet with $J=J_2$ and at $J_3\approx0.65$ in the other antiferromagnet for all wave vectors. For $J_3\approx J_2$ even in the near-boundary layers the spectrum consists of a continuum of bulk states [see Fig.~\ref{fig1}(b)]. New peaks start to form near the upper edges of the continuum spectra at $J_3\approx J_1$ in the antiferromagnet with $J=J_2$, and at $J_3\approx1.2$ in the second antiferromagnet [see Fig.~\ref{fig1}(c)]. As follows from Fig. 3, these peaks are seen only in the interface region and, therefore, they correspond to near-boundary modes. For moderate $J_3$ only one of the two peaks and its replica in the neighbor antiferromagnet may be observed for some momenta $\e k$ [see Figs.~\ref{fig3}(a) and (c)]. For other $\e k$ both near-boundary peaks and their replicas are seen [Figs.~\ref{fig3}(b) and (d)]. In the antiferromagnet with $J=J_1$ the dispersion of the mode resembles in shape the dispersion of the 2D spin waves, however, with a finite offset of frequencies by $\omega_0\approx3J_1$,
\begin{equation}
	\omega_{\e k}=\omega_0+2J'\sqrt{1-\left(\gamma^{(2)}_{\e k}\right)^2}.
\end{equation}
For parameters of Fig. 3 $J'\approx0.13$, much smaller than $J_1$.
In this respect this mode differs from the near-boundary modes existing at $J_3\ll J_2$, which dispersions correspond to effective exchange constants larger than $J_1$ (see Fig.~\ref{fig2}). Apparently in the limit of very large $J_3$ the near-boundary modes in Fig.~\ref{fig3} transform to triplet excitations of the pairs of boundary spins.

\section{Spin correlations}
Spin correlations between nearest neighbors can be expressed in terms of correlations of spin-wave operators using Eq.~\eqref{2} and the translational invariance of Hamiltonian \eqref{3} in the $YZ$ plane,
\begin{eqnarray}\label{8}\nonumber
    \langle \e S_{\e L}\e S_{\e L'}\rangle=
    \frac{1}{2N}\sum_{\e k}\biggl\{2\cos[\e k(\e l-\e l')]\langle b_{\e k l_x}b_{-\e k,l'_x}\rangle\\
    +\langle b^\dag_{\e k l_x}b_{\e k l_x}\rangle+\langle b^\dag_{\e k l'_x}b_{\e k l'_x}\rangle\biggr\}-\frac14.
\end{eqnarray}
Bearing in mind the property of Green's function \eqref{6}
\begin{equation*}
    D_{ij}(\e k \omega l_x l'_x)=D_{ji}(\e k \omega l'_x l_x),
\end{equation*}
the spin-wave correlations in Eq. \eqref{8} can be expressed as
\begin{equation}\label{9}
    \left\langle \hat B_{\e k l_x}\hat B^\dag_{\e k l'_x}\right\rangle=
    \int_{-\infty}^\infty \frac{d\omega}{\pi}\frac{\mathrm{Im}[\hat D(\e k \omega l_x l'_x)]}{e^{-\omega\beta}-1},
\end{equation}
where $\beta=T^{-1}$ is the inverse temperature.

Let us denote the spin correlations on sites located parallel and perpendicular to the interface as
\begin{equation*}\label{10}
\begin{split}
    C_L(l_x)=\left\langle \e S_{\e l+\e a,l_x}\e S_{\e l l_x}\right\rangle,\\
    C_T\left(l_x+\frac12\right)=\left\langle \e S_{\e l,l_x+1}\e S_{\e l l_x}\right\rangle.
\end{split}
\end{equation*}
\begin{figure}[h]
\begin{center}
\includegraphics[width=\linewidth]
{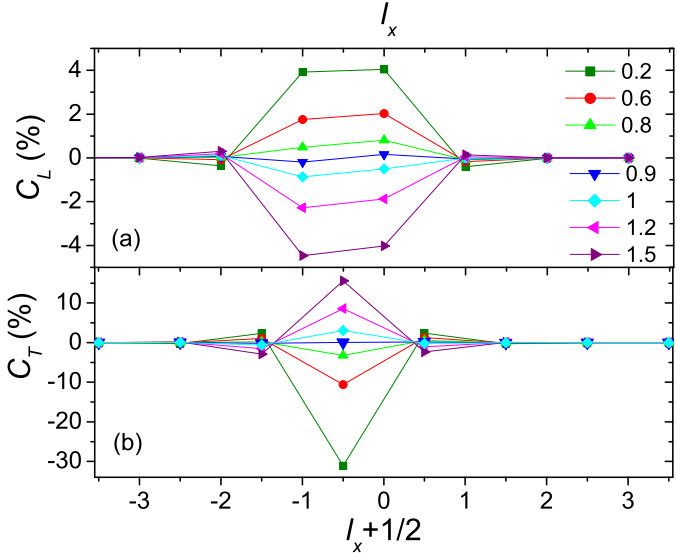} \caption{The deviations of the nearest-neighbor spin correlations from the bulk value as functions of $l_x$ for sites located parallel ($\Delta C_L $, a) and perpendicular ($\Delta C_T $, b) to the interface. $J_2 =0.8$, values of $J_3$ are indicated in the figure.} \label{fig4}
\end{center}
\end{figure}
For $T=0$ with the use of Eq.~\eqref{9} these spin correlations read
\begin{eqnarray}
    C_L(l_x)&=&-\frac{1}{N}\sum_{\e k} \cos(k_y)\int\limits_{0}^{\infty}\frac{d\omega}{\pi}\mathrm{Im} D_{12}(\e k \omega l_x l_x)\nonumber\\
    &-&\frac{1}{N}\sum_{\e k}\int\limits_{0}^{\infty}\frac{d\omega}{\pi}\mathrm{Im}D_{22}(\e k \omega l_x l_x)-\frac14,\nonumber
\end{eqnarray}
\begin{eqnarray}
    &&\hspace{-0.6cm}C_T\!\left(l_x+\frac12\right)=\frac{1}{2N}\sum_{\e k}\int\limits_0^{\infty}\frac{d\omega}{\pi}\,\mathrm{Im} D_{22}(\e k \omega l_x l_x)\\
    &&\hspace{1cm}-\frac{1}{2N}\sum_{\e k}\int\limits_0^{\infty}\frac{d\omega}{\pi} \,\mathrm{Im}D_{22}(\e k \omega,l_x+1,l_x+1)\nonumber\\
    &&\hspace{1cm}-\frac{1}{N}\sum_{\e k}\int\limits_0^{\infty}\frac{d\omega}{\pi}\,\mathrm{Im}D_{12}(\e k \omega l_x,l_x+1)-\frac14.\nonumber
\end{eqnarray}
Let us define deviations of spin correlations parallel and perpendicular to the interface from the bulk spin correlation $C_b=-0.3005$ \cite{voropajeva2009,oitmaa} as
\begin{equation*}
    \Delta C_{L,T}=\frac{C_{L,T}-C_b}{C_b}.
\end{equation*}

\begin{figure}[h]
\begin{center}
\includegraphics[width=\linewidth]
{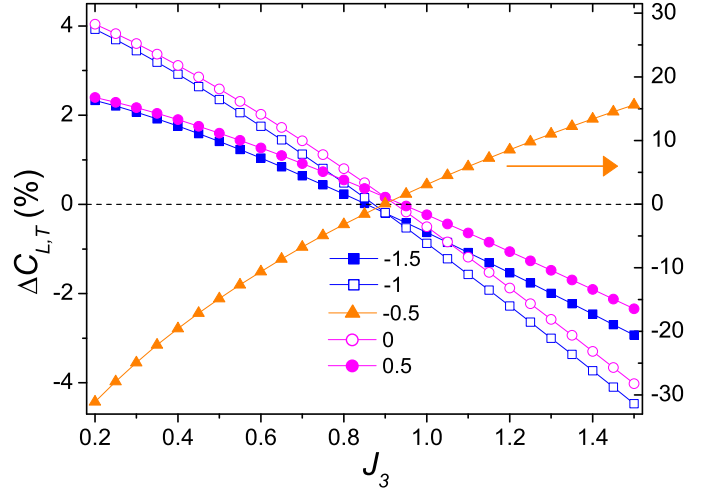} \caption{The deviations $\Delta C_{L,T} $ as functions of $J_3 $ for different values of $l_x $ and $l_x+1/2$ indicated in the figure. The right $y$ axis corresponds to $\Delta C_T(-0.5)$, the left $y$ axis to all other quantities.} \label{fig5}
\end{center}
\end{figure}

The calculated values of these deviations are shown in Figs.~\ref{fig4}  and~\ref{fig5} for $T=0$. As seen from these figures, main deviations of the spin correlations from the bulk value fall on the boundary and the second to the boundary layers, i.e. on the existence domain of the near-boundary modes. The largest in absolute value deviations are obtained in and between the boundary layers for $J_3\ll J_2$ and for $J_3>J_1$, when the near-boundary modes are well resolved in the spectral function. As indicated in Refs.\ \cite{voropajeva2009,voropajeva2010,sherman}, in the case of a semi-infinite antiferromagnet the amplified spin correlations in the boundary layer are connected with the quasi-two-dimensionality of the near-boundary mode, which dominates in the spectral function of this layer. It behaves like a 2D antiferromagnet, in which the nearest-neighbor spin correlation is larger in modulus than in a three-dimensional antiferromagnet (0.3346 vs.\ 0.3005 \cite{gochev,oitmaa}). This explains the strengthened spin correlations in this layer and between it and the second to the boundary layer. The correlations in the second and between the second and third layers are smaller in modulus than $C_b$ due to the destructive contribution of the boundary and bulk modes. For the case $J_3\ll J_2$ in both antiferromagnets the distribution of spin correlations is similar to that mentioned above and apparently has the same explanation. Peaks of the near-boundary mode disappear at $J_3\approx0.5$ in the antiferromagnet with $J=J_2$ and at $J_3\approx0.65$ in the other antiferromagnet, which allows bulk modes to penetrate in the near-boundary region. As a consequence spin correlations in the second and between the second and third layers of an antiferromagnet become approximately equal to $C_b$. The new type of interface excitations appears at $J_3\approx J_1$ in the antiferromagnet with $J=J_2$ and at $J_3\approx1.2$ in the other antiferromagnet. In this mode, spin correlations between sites on each side of the interface and in the second and between the second and the third layers are amplified, while correlations on the boundary and between the boundary and the second layers are weakened in both antiferromagnets. As follows from Figs.~\ref{fig4} and~\ref{fig5}, the deviations of the spin correlations from the bulk value change sign at $J_3\approx(J_1+J_2)/2$.

\section{Conclusion}

In this Letter, magnetic excitations and nearest-neighbor spin correlations near the interface of two semi-infinite spin-$\frac12$ Heisenberg antiferromagnets were investigated using the spin-wave approximation. It was assumed that the antiferromagnets have identical simple cubic lattices, the nearest-neighbor exchange constants $J_1$ and $J_2=0.8J_1$ in the bulk of the crystals and $J_3$ between the interface sites. It was shown that for $J_3 \ll J_2$ a near-boundary region arises, which embraces four layers closest to the interface. In this region, magnetic excitations are quasi-two-dimensional spin waves, which expel bulk modes from their domain. As $J_3$ approaches $J_2$ peaks of the near-boundary modes lose intensity and finally disappear in the continuum of the bulk modes, which penetrate in the near-boundary region. This process starts in the antiferromagnet with the smaller exchange constant and at small wave vectors. At $J_3\approx J_1$ in the antiferromagnet with $J=J_2$ and then, at $J_3\approx1.2J_1$ in the other antiferromagnet a new type of near-boundary modes arises. In these modes pairs of spins on each side of the interface are mainly involved. The separation of the heterostructure into the regions with different spin excitations leads to a peculiar distribution of nearest-neighbor spin correlations near the interface. In the case $J_3\ll J_2$ the decreased dimensionality of the near-boundary modes leads to the correlations in the boundary layers, which are increased in modulus in comparison with the bulk correlation. At the same time the destructive contribution of the near-boundary and bulk modes decreases correlations between some sites. For $J_3\approx (J_1+J_2)/2$ the correlations are close to the bulk one. In the case  $J_3>J_1$ the correlations are amplified between the interface spins, and they are weakened in the boundary layers of both antiferromagnets. The deviations of the correlations from the bulk value reverse sign at $J_3\approx (J_1+J_2)/2$.

\section*{Acknowledgements}
This work was supported by the European Regional Development Fund (Centre of Excellence "Mesosystems: Theory and applications", TK114) and by the ETF grant No.~9371.

\end{document}